\documentclass[10pt,superscriptaddress,twocolumn,amsmath,amssymb,aps,prl,showpacs]{revtex4-1}
\usepackage{mathrsfs}
\usepackage{array}
\usepackage{graphicx}
\usepackage{dcolumn}
\usepackage{bm}
\usepackage[title,titletoc,header]{appendix}
\usepackage{amssymb}
\usepackage{amsmath}
\usepackage{mathrsfs}
\usepackage{amsmath}
\usepackage{subfigure}
\usepackage{float}
\usepackage[title,titletoc,header]{appendix}
\usepackage{graphicx}
\usepackage{color}

\newcommand{\be}{\begin{equation}}
\newcommand{\ee}{\end{equation}}
\newcommand{\bea}{\begin{eqnarray}}
\newcommand{\eea}{\end{eqnarray}}

\begin{document}
\title{Superfluid Density of a Spin-orbit Coupled Bose Gas}
\author{Yi-Cai Zhang}
\affiliation{Department of Physics and Center of Theoretical and
Computational Physics, The University of Hong Kong, Hong Kong, China}

\author{Zeng-Qiang Yu}
\affiliation{Institute of Theoretical Physics, Shanxi University, Taiyuan 030006, China}

\author{Tai Kai Ng}
\affiliation{Department of Physics, Hong Kong University of Science and Technology,
Kowloon, Hong Kong, China}

\author{Shizhong Zhang}
\email[]{shizhong@hku.hk}
\affiliation{Department of Physics and Center of Theoretical and
Computational Physics, The University of Hong Kong, Hong Kong, China}

\author{Lev Pitaevskii}
\affiliation{INO-CNR BEC Center and Department of Physics, University of Trento, Via Sommarive 14, Povo, Italy}
\affiliation{Kapitza Institute for Physical Problems RAS, Kosygina 2, 119334 Moscow, Russia}

\author{Sandro Stringari}
\affiliation{INO-CNR BEC Center and Department of Physics, University of Trento, Via Sommarive 14, Povo, Italy}

\date{\today}
\begin{abstract}
We discuss the superfluid properties of a Bose-Einstein condensed  gas with spin-orbit coupling, recently realized in experiments. We find a finite normal fluid density $\rho_n$ at zero temperature which turns out to be a  function of the Raman coupling. In particular, the entire fluid becomes normal at the transition point from the zero momentum to the plane wave phase, even though the condensate fraction remains finite. We emphasize the crucial role played by the gapped branch of the elementary excitations and  discuss its contributions to various sum rules. Finally, we prove that an independent definition of superfluid density $\rho_s$, using the phase twist method,  satisfies the  equality $\rho_n+\rho_s=\rho$, the total density, despite the breaking of Galilean invariance. 
\end{abstract}

\maketitle

{\em Introduction}. Bose-Einstein condensation and superfluidity are two distinct but intimately related phenomena~\cite{Leggett2006, LS2003}. Usually, the existence of a finite condensate fraction implies that the system should behave as a superfluid, characterized by a non-zero superfluid density $\rho_s$. This is the case, for example, of superfluid $^4$He where the condensate fraction is about $10\%$,  while the superfluid density coincides with the total density ($\rho_s=\rho$) at zero temperature.  A convenient quantity to define  is the so-called normal density $\rho_n$, which quantifies the amount of the fluid in equilibrium with a moving wall~\cite{Baym1969}. For system with Galilean invariance, it can be shown that $\rho_n+\rho_s=\rho$. As a result, for $^4$He at zero temperature, $\rho_n=0$. The vanishing of the normal density in $^4$He is related to the scarcity of long wave length excitations, for which only phonons are available~\cite{GN1964}. 

Recently, a new type of Bose-Einstein condensate has been realized in ultracold atomic gases with synthetic spin-orbit coupling~\cite{Lin2011,JYZhang,JWPan,Zwierlein,JZhang1,JZhang2,JZhang3,Hamner,Olson,SCJ,Engels2014}. Two novel features stand out in comparison with the usual Bose-Einstein condensates. Firstly, the Galilean invariance of this novel system is broken due to spin-orbit coupling~\cite{Zhu2012,Zheng2013,Ozawa2013}. This has important consequences on the behavior  of the dipolar oscillation in  a harmonic trap~\cite{JYZhang},  and in particular on its hybridized density/magnetic nature~\cite{Yun2012b}. Secondly, in spin-orbit coupled BECs, the long wave length excitations basically consist of two branches~\cite{Martone2012}: a phononic excitation with linear dispersion like that in $^4$He, and a gapped branch dominated by spin excitations, in accordance with experiments~\cite{SCJ,Engels2014}. The new structure of elementary excitations   and the breaking of Galilean invariance are expected to  introduce new effects in the   superfluid properties of spin-orbit coupled systems (for recent reviews see, for example, Ref.~\cite{YunLi2015,Hui2015}). 

In this Letter we show that, for a spin-$1/2$ boson subjected to spin-orbit coupling with equal superposition of the Rashba and Dresselhaus terms, the normal density $\rho_n$ remains finite even at zero temperature. This is an important consequence of the breaking of  Galilean invariance. An analogous situation takes place in the presence of disorder or of an external periodic potential. With respect to these latter cases, the spin-orbit Hamiltonian employed in the present work has the peculiarity of being translational invariant.  We study how $\rho_n$ evolves with the Raman coupling and show that the entire fluid becomes normal at the transition point from the plane wave phase to the zero momentum phase, despite the presence of Bose-Einstein condensation. We show that the existence of the gapped branch in the long wave length excitation is responsible for the finite value of the normal density  and its role in the calculation of various sum rules is discussed. Finally, we give an independent definition of the superfluid density using the phase twist method and show that, despite the lack of Galilean invariance, the equality $\rho_s+\rho_n=\rho$ still holds. 

{\em Spin-orbit coupled Bose condensate}. The single particle Hamiltonian of a spin-orbit coupled Bose gas is given by (for simplicity, we set $\hbar=m=1$)~\cite{Ho2011,Yun2012a,Yun2013,Yu2014}
\be
h_0=\frac{1}{2}[(p_x-k_0 \sigma_z)^2+p_y^2+p^2_z]+\frac{\Omega}{2}\sigma_x+\frac{\delta}{2}\sigma_z,
\label{h0}
\ee
where $k_0$ is the momentum transfer from the two Raman lasers, which we assume to be oriented along the $\hat{x}$-direction and ${\bf p} =-i\nabla$ is the canonical momentum, not to be confused with the physical momentum whose $x$-component reads $P_x= p_x-k_0 \sigma_z$. The quantity $\Omega$ is the two-photon Rabi frequency determined by the intensity of the Raman lasers and $\delta$ is the Raman detuning, which we will  set equal to zero in the following discussions. Finally  the operators $\boldsymbol{\sigma}=(\sigma_x,\sigma_y,\sigma_z)$ are the usual Pauli matrices describing the two internal states of the atoms. The single-particle Hamiltonian $h_0$ is translational invariant ($[p_x, h_0]=0$), but  breaks Galilean invariance since it does not commute with the physical momentum ($[P_x,h_0]\ne 0$).

For $^{87}$Rb atoms employed in current experiments, the interaction between atoms can be written as $V_{\rm int}=1/2\sum_{\alpha\beta}\int d{\bf r} g_{\alpha\beta}n_\alpha({\bf r})n_\beta({\bf r})$, where $g_{\alpha\beta}=4\pi a_{\alpha\beta}$ are the various coupling constants in different spin channels, with $a_{\alpha\beta}$  the corresponding scattering lengths ($\alpha, \beta=\uparrow,\downarrow$ label the relevant internal hyperfine-Zeeman states). In the following we will assume   $g_{\uparrow\uparrow}=g_{\downarrow\downarrow}\equiv g$, while $g_{\uparrow\downarrow}=g_{\downarrow\uparrow}\equiv g'$ is  not necessarily equal to $g$. In the present work, we shall only consider  uniform gases in the absence of external harmonic traps. 

The Hamiltonian (\ref{h0}) has been implemented experimentally and both the phase diagram~\cite{JWPan} and the elementary excitations~\cite{SCJ} have been investigated. Following Ref.~\cite{Yun2012a}, let us define the interaction paramters $G_1=n(g+g')/4$ and $G_2=n(g-g')/4$, where $n=N/V$ is the average density. Then one can predict three different quantum phases. For small Rabi frequency $\Omega$ and $G_2>0$, a stripe phase with density modulation in the ground state exists. The low frequency elementary excitations of the stripe phase consist of two gapless modes associated with the spontaneous breaking of translational and gauge symmetries. For relatively larger values of $\Omega$, two new phases emerge where the condensate wave function can be written in the following form
\be
\left|0\right\rangle=\sqrt{n}\left[\begin{array}{c} \cos\theta\\ -\sin\theta\end{array}\right]\exp(ik_1x).
\ee
Minimizing the mean field energy, one finds that, for $\Omega<2(k_0^2-2G_2)$, the ground state configuration is characterized by  $k_1=k_0\sqrt{1-\Omega^2/[2(k_0^2-2G_2)]^2}$ and $\cos2\theta=k_1/k_0$. This state breaks the $Z_2$ symmetry of the Hamiltonian and features a non-zero magnetization in the ground state, given by $M/N\equiv \langle \sigma_z\rangle= k_1/k_0$.  This phase is usually referred to as the plane wave phase. For $\Omega>2(k_0^2-2G_2)$, one has $k_1=0$ and $\theta=\pi/4$. This gives rise to the zero momentum phase. The phase transition between these two phases is of second order nature with a divergent thermodynamic magnetic susceptibility $\chi_M$ at the critical point~\cite{Yun2012b}. The elementary excitations of these two phases consist of two branches:  a gapless phonon branch $\omega_1({\bf q})$ corresponding, in the small $q$ limit,  to the sound propagation and a gapped branch $\omega_2({\bf q})$ dominated by spin excitations. The sound velocity along the $\hat{x}$-direction is strongly quenched by spin-orbit coupling  and phonons exhibit  a hybridized density/spin nature \cite{Martone2012}. In fact, at the transition point, taking place at the Rabi frequency  $\Omega_c\equiv2(k_0^2-2G_2)$, the sound velocity vanishes, even though the compressibility of the gas remains finite. This  implies that the Landau's critical velocity along $\hat{x}$-direction becomes zero, according to the usual Landau criterion. On the other hand, the condensate density remains finite, the quantum depletion being small even at the transition point~\cite{Zheng2013}.

{\em Normal density}. According to the usual concept of two-fluid hydrodynamics, the normal density is the  fraction of the fluid which is    dragged by the wall of a moving cylindrical container, while the superfluid component can move with respect to the container without friction. In our case the role of the wall is played by the Raman lasers which block the motion of the normal component of the gas along the $\hat{x}$-direction.  As a result, the normal density is a tensor of the form $\hat{\rho}_n=\rho_{n,\parallel}\hat{x}\hat{x}+\rho_{n,\perp}(\hat{y}\hat{y}+\hat{z}\hat{z})$. The transverse component $\rho_{n,\perp}$ behaves as usual since excitations along the transverse directions remain the same in the long wave length limit, thereby yielding $\rho_{n,\perp}=0$ at zero temperature. On the other hand, the longitudinal component $\rho_{n,\parallel}$  (hereafter denoted as $\rho_n$) is modified significantly due to spin-orbit coupling.

In terms of the transverse current response function~\cite{Baym1969}, the normal density $\rho_n$ at zero temperature is given by~\cite{notesymm}
\begin{equation}
\frac{\rho_n}{\rho}=\frac{1}{N}\lim_{{\bf q}\to 0}\left[\sum_{n\ne 0}\frac{|\langle0|J_x^{T}({\bf q})|n\rangle|^2}{E_n-E_0}+ ({\bf q}\to-{\bf q})\right]
\label{defrhon}
\end{equation}
where  $J_x^{T}({\bf q})$ is the transverse current operator along the  $\hat{x}$-direction (${\bf q}\perp \hat{x}$). Here $|n\rangle$ is the set of exact many-body eigenstate with energy $E_n$. We take ${\bf q}=q\hat{y}$ so that the transverse current operator takes  the explicit form 
\begin{equation}
J_x^{T}(q)=\sum_k (p_{k,x} - k_0 \sigma_{k,z})e^{iqy_k},
\label{JT}
\end{equation}
where $k$ enumerates the number of particles. Since the transverse current operator does not excite the gapless phonon mode, which is of longitudinal nature, the only contribution to Eq.(\ref{defrhon}) comes from the gapped  branch. Let us denote the gap in the limit $q\to 0$ as $\Delta\equiv\omega_2(q=0)$, then we can write Eq.(\ref{defrhon}) as
\begin{align}
&\frac{\rho_n}{\rho} =\frac{1}{N\Delta^2}\lim_{q\to  0}\left[\sum_{n \ne 0}|\langle 0|J_x^{T}(q)|n\rangle|^2(E_n-E_0) + (q \to -q)\right] \notag \\
&=\frac{1}{N\Delta^2}\lim_{q\to 0}\langle 0|[J_x^{T}(-q),[H,J_x^{T}(q)]]|0\rangle.
\end{align}
In the limit $q\to 0$,  the only non-commutating term between the current operator and the Hamiltonian is the spin term; the contribution of the canonical component of the current vanishing when $q\to 0$, as a consequence of the translational invariance of the Hamiltonian. One can consequently write $\lim_{q\to 0}\langle 0|[J_x^{T}(-q),[H,J_x^{T}(q)]]|0\rangle   
=k_0^2\langle 0|[\Sigma_z,[H,\Sigma_z]]|0\rangle$ where $\Sigma_z= \sum_k\sigma_{k,z}$ is the total spin operator along the ${\hat z}$-direction. 
 The double commutator only receives contribution from the Raman term proportional to $\Omega$ in the single-particle Hamiltonian, yielding  $[\Sigma_z,[H,\Sigma_z]]=-2N\Omega \sigma_x $. One finally obtains  the  result 
\begin{equation}
\frac{\rho_n}{\rho}=-\frac{2k_0^2\Omega}{\Delta^2}\langle \sigma_x\rangle.
\label{rhon1}
\end{equation}

A further important connection between the superfluid and the magnetic properties of the system is given by the nontrivial  identity  
\begin{equation}
\frac{\rho_n}{\rho_s}=k_0^2\chi_M,
\label{mainresult}
\end{equation}
for the ratio between the normal density ($\rho_n$) and the superfluid ($\rho_s= \rho- \rho_n$) density of the system. The quantity $\chi_M$ entering Eq.(\ref{mainresult}) is the thermodynamic magnetic susceptibility of the system, determined by the energy cost $\delta E=N(\delta \langle \sigma_z \rangle)^2/(2\chi_M)$ associated with the change $\delta\langle \sigma_z \rangle $ in the polarization of the medium. In the following we will show that the identity  (\ref{mainresult}) holds both in  the plane wave and in the zero momentum  phase.

{\em (i) Plane wave phase} ($\Omega \le \Omega_c=2(k^2_0-2G_2)$). The transverse spin polarization is given by $\langle\sigma_x\rangle=-\Omega/2(k_0^2-2G_2)$ and the excitation gap is given by $\Delta^2=4(k_0^2-2G_2)(k_0^2-2G_2k_1^2/k_0^2)$~\cite{Martone2012}. As a result, we have
\begin{equation}
\frac{\rho_n}{\rho}=\frac{k_0^2\Omega^2}{4(k_0^2-2G_2)^3+2G_2\Omega^2}.
\label{main2}
\end{equation}
In the plane wave phase $\chi_M=\Omega^2/\{(k_0^2-2G_2)[4(k_0^2-2G_2)^2-\Omega^2)]\}$~\cite{Martone2012}, thus confirming the relation (\ref{mainresult}). 

{\em (ii) Zero momentum phase} ($\Omega \ge \Omega_c$). The transverse spin polarization is given by $\langle\sigma_x\rangle=-1$ and  the excitation gap is given by $\Delta^2=\Omega(\Omega+4G_2)$~\cite{Martone2012}. Thus we find
\begin{equation}
\frac{\rho_n}{\rho}=\frac{2k_0^2}{\Omega+4G_2}.
\label{main1}
\end{equation}
In the zero momentum phase, $\chi_M=2/(\Omega+4G_2-2k_0^2)$~\cite{Martone2012}, thus confirming again the relation (\ref{mainresult}). 

\begin{figure}
\begin{center}
\includegraphics[width=0.6\columnwidth]{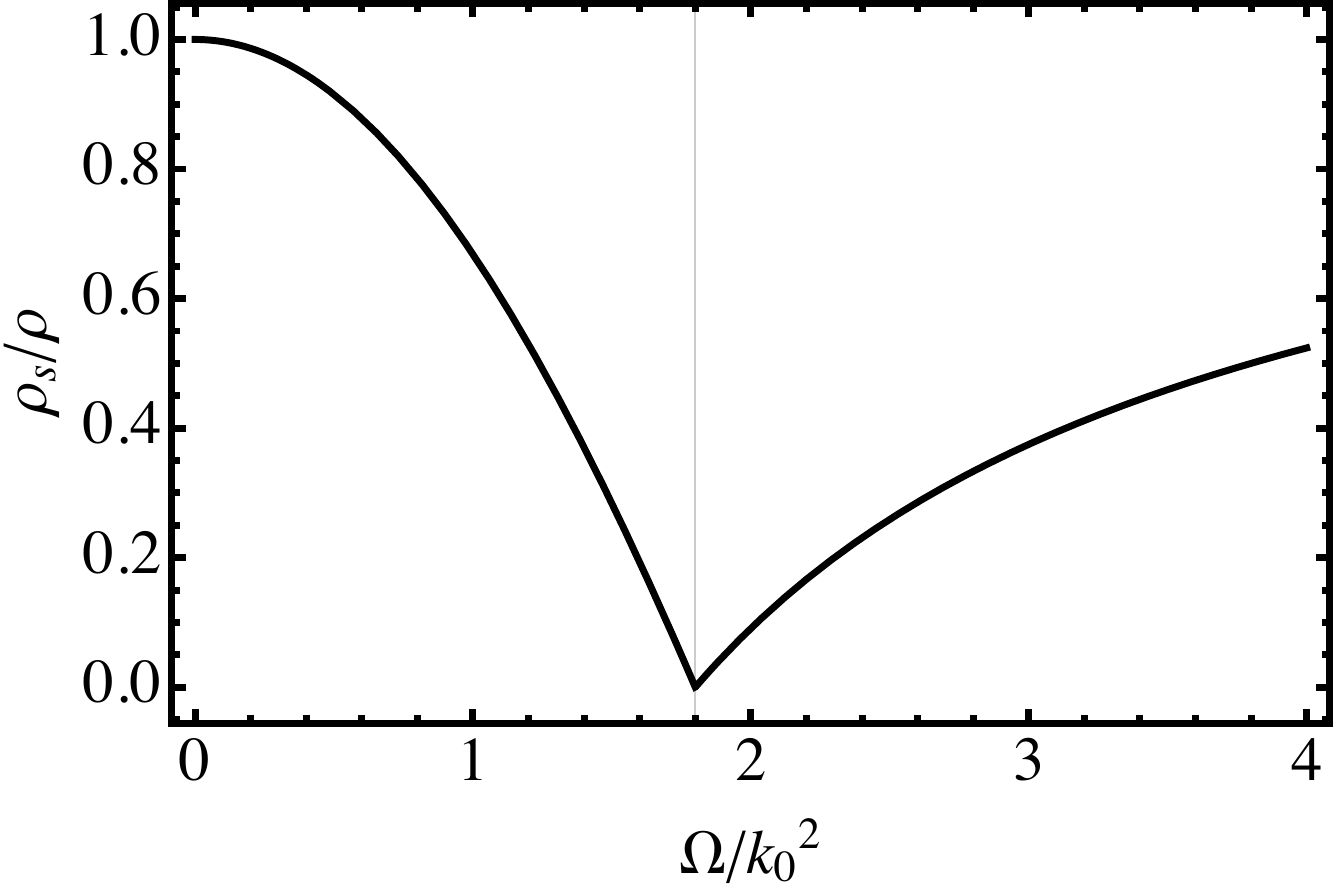}
\end{center}
\caption{Dependence of $\rho_s/\rho$ on the Raman coupling strength $\Omega$ at $T=0$ ($G_1=0.2k_0^2$, $G_2=0.05k_0^2$). The figure reveals the strong quenching of the superfluid density, which exactly vanishes at $\Omega_c$. }
\label{figrhos}
\end{figure}

Eqs. (\ref{main2},\ref{main1}) show that the normal density takes the maximum value at the transition point ($\Omega=\Omega_c$) between the two phases, where $\rho_n/\rho=1$ and hence $\rho_s=0$, consistent with the divergent behavior of $\chi_M$~\cite{Martone2012}. At $\Omega=\Omega_c$, the entire fluid becomes normal even though the condensate fraction is finite~\cite{Zheng2013}. The explicit dependence of $\rho_s/\rho$ on the Raman coupling $\Omega$ is shown in Fig.\ref{figrhos}.

In both the zero momentum and the plane wave phases, the normal density depends explicitly  on the interaction parameters $G_2$, which quantifies the breaking of the $SU(2)$ invariance of the inter-atomic force. In the limit $G_2=0$, one finds $\rho_n/\rho=\Omega^2/4k_0^4$ for the plane wave phase and $\rho_n/\rho=2k_0^2/\Omega$ for the zero momentum phase. These results can also be written in the useful form
\begin{align}
\frac{\rho_n}{\rho}=1-\frac{m}{m^*},
\end{align}
where $m^*$ is the effective mass of atoms close to the single particle energy minimum of the Hamiltonian (\ref{h0}), given by $m/m^*= 1- (\Omega/2k^2_0)^2$ in the plane wave phase and $m/m^*=1-2k^2_0/\Omega$ in the zero momentum phase~\cite{Zheng2013,YunLi2015}. These results show that if the interaction is $SU(2)$ invariant, the normal density in the ground state is independent of inter-particle interaction and is controlled entirely by the Raman lasers. 

{\em Sum-rule analysis}. It is by now clear that the existence of the gapped branch in the elementary excitation spectrum is responsible for the finite normal density even at zero temperature. To gain further insight into the problem, we investigate the   moments~\cite{LS2003}
\begin{equation}
m_p=\int d\omega \omega^p S({\bf q},\omega)= \sum_n(E_n-E_0)^p|\langle n|\rho_{\bf q}|0\rangle|^2
\label{sumrules}
\end{equation}
of the density dynamical structure factor $S({\bf q},\omega)$, where $\rho_{\bf q}$ is the density fluctuation operator. It can be verified directly that, even in the presence of spin-orbit coupling, the $p=1$ moment obeys the well known $f$-sum rule: 
\begin{equation}
m_1({\bf q})+m_1(-{\bf q}) =  \langle 0|[[\rho_{\bf q},H],\rho_{\bf q}^\dagger]|0\rangle=Nq^2 \; .
\label{fsumrule}
\end{equation}
Using the continuity equation $[\rho_{\bf q},H]= \omega\rho_{\bf q}={\bf q}\cdot{\bf J}^L({\bf q})$, where $J_x^{L}({\bf q})=\sum_k(p_{k,x}e^{iqx_k} +e^{iqx_k}p_{k,x})/2 - k_0\sum_k \sigma_{k,z}e^{iqx_k} $ is the longitudinal current operator with   ${\bf q}=q\hat{x}$, we can rewrite the $f$-sum rule Eq.(\ref{fsumrule}) in terms of matix elements of ${\bf J}^L({\bf q})$ as
\begin{align}
Nq^2=\sum_{n\ne 0}\frac{|\langle n|J_x^L({\bf q})|0\rangle|^2}{E_n-E_0}q^2 + (q\to -q),
\label{fsumrule2}
\end{align}
where the summation over $n$ includes both the phonon and the gapped branch. For the phonon branch $(E_n-E_0)\sim cq$ as $q\to 0$ with $c$ being the sound velocity. The corresponding matrix element $\langle n|J^L_x({\bf q})|0\rangle$ vanishes like  $\sqrt{q}$. On the other hand, for the gapped branch, $(E_n-E_0)\to \Delta$ as $q\to 0$ and the matrix element $\langle n|J^L({\bf q})|0\rangle$ approaches a constant value since the operator $P_{x}=\sum_k(p_{k,x}-k_0\sigma_{k,z})$  does not commute with the Hamiltonian. As a result, both the gapped and the phonon branch contribute to the $f$-sum rule in the long wave length limit. This differs from  the usual situation (see, for example, liquid $^4$He) where the phonon contribution dominates at small values of $q$ and exhausts the $f$-sum rule. Actually, the 
contributions $|\langle n|J^T_x({\bf q})|0\rangle|^2$ and $|\langle n|J^L_x({\bf q})|0\rangle|^2$ arising from the gapped branch and entering the transverse and longitudinal sume rules  (\ref{defrhon}) and (\ref{fsumrule}), coincide in the $q\to 0$ limit. Consequently the normal density fraction $\rho_n/\rho$ is fixed by the contribution of the gapped branch to the $f$-sum rule. 

Analogously, we can investigate the contributions arising from the phonon and from the gapped branch to the other sum rules. In the $q\to 0$ limit,  the phonon contribution is of order $q^{p+1}$, while for the gapped branch, it is always $q^2$. As a result, the dominant contribution to the   compressibility sum rule $m_{-1}$ and to the static structure factor $m_0$ arises from the phonon branch. The higher order sum rules $m_p$ with $p>1$ are instead exhausted by the gapped branch.

An important consequence of the above analysis is that the superfluid density along the $\hat{x}$-direction corresponds to the phonon contribution to the inverse energy weighted sum rule (\ref{fsumrule2}) relative to the longitudinal current operator. On the other hand, using  the continuity equation $qJ_x(\pm q)=\omega_\pm(q)\rho_q$, where $\omega_\pm(q)$ labels the excitation frequency along the positive (negative) $\hat{x}$-direction with sound velocity $c_+$ ($c_-$): $\omega_\pm(q)=c_\pm q$ \cite{c+c-}, and   the fact that the $m_{-1}$ sum rule is exhausted by the phonon branch, we can write \cite{note10}
\begin{equation}
\rho_s=\rho c_-c_+\kappa,
\label{c+c-}
\end{equation}
where $\kappa$ is the thermodynamic compressibility. This equation shows that  the measurements of the sound velocities along the $\pm\hat{x}$-direction and the knowledge of the static compressibility are enough for a direct determination of the superfluid density. In the absence of spin-orbit coupling, $\rho_s=\rho$ at zero temperature and Eq.(\ref{c+c-}) reduces to the standard relation $\kappa c^2=1$. Since the compressibility of the gas is practically unaffected by the spin-orbit coupling~\cite{YunLi2015}, the measured quenching of the sound velocities (see Figure 3 in~\cite{SCJ}), provides  direct evidence for  the suppression of the superfluid density near the transition, as revealed by Figure 1. According to Eq. (\ref{c+c-}), the sound propagating along the $\hat{x}$-direction with velocities $c_{\pm}$ can then be regarded as the fourth sound~\cite{Atkins1959}, characterized by the motion of the superfluid component, while the normal component $\rho_n$ remains at rest.

The above discussion has emphasized the deep difference, typical of superfluids,  between  the longitudinal and the transverse current   response functions. In particular   the transverse static response [Eq.(\ref{defrhon})], differently from the longitudinal one [Eq.(\ref{fsumrule})], is proportional to the normal (non superfluid) component of the gas. At high frequencies, however, both response functions instead exhibit,  when $q\to 0$,  the leading $1/\omega^2$-dependence with the coefficients given by 
\begin{equation}
\lim_{q\to  0}\sum_n|\langle 0|J_x^{T,L}(q)|n\rangle|^2(E_n-E_0)=\frac{k^2_0}{2} \langle 0|[\Sigma_z,[H, \Sigma_z]]|0\rangle,
\end{equation}
determined by the energy weighted sum rule $\langle 0|[\Sigma_z,[H, \Sigma_z]]|0\rangle= -2N\Omega \langle \sigma_x\rangle$.

{\em Proof of the identity $\rho_n+\rho_s=\rho$}. The superfluid density  can be defined microscopically by employing the phase twist method~\cite{Fisher1973}. The phase twist  can be generated  by  the unitary transformation (Galilean boost) 
\begin{equation}
|\Psi'\rangle=\exp(i\theta\sum_kx_k/L_x)|\Psi\rangle \; ,
\label{unitary}
\end{equation}
applied to the wave function $|\Psi\rangle$ where $L_x$ is the length of the system and $|\Psi\rangle$ obeys the usual periodic boundary conditions. The many-body wave function $|\Psi'\rangle$ is then characterized by a   phase twist $\varphi(L_x)-\varphi(0)= \theta$.  At $T=0$, minimization of the energy for a fixed and small value of $\theta$   defines the superfluid density $\rho_s$ according to 
\begin{equation}
E'-E\equiv\frac{1}{2}N\frac{\rho_s}{\rho}\left(\frac{\theta}{L_x}\right)^2 \; ,
\end{equation} 
where    $E'$ and $E$  are the ground state energies in the presence and in the absence of the twist constraint, respectively. According to  (\ref{unitary}) the physical momentum operator $P_x$ acts on $|\Psi'\rangle$  as $ P_x|\Psi'\rangle=\exp(i\theta\sum_kx_k/L_x)( P_x+ N\theta/L_x)|\Psi\rangle$ and consequently  
the calculation of $E'$  corresponds to minimizing the energy with respect to $|\Psi \rangle$ with a modified Hamiltonian:
\begin{equation}
\langle \Psi'|H  |\Psi' \rangle= \langle \Psi|\left[H + \frac{1}{2}\left(\frac{\theta}{L_x}\right)^2 N+\frac{\theta}{L_x} P_x\right] |\Psi \rangle.
\end{equation}

The energy difference $E'-E$, and hence $\rho_s$,  can be easily calculated by second order perturbation theory. Noting that $\langle 0|P_x|0\rangle=0$, since there is no net current in the ground state, we find that the superfluid density is eventually given by 
\begin{equation}
\rho_s=\rho\left(1-\frac{2}{N}\sum_{n\neq 0}\frac{|\langle 0|P_x|n\rangle|^2}{E_n-E_0}\right).
\end{equation}
Since only the spin component of the operator $P_x$ gives rise to non vanishing matrix elements (and in particular only the upper branch can be excited), one finds that the sum entering the above equation coincides with Eq.(\ref{defrhon}) in the small $q$ limit, which defines the normal density. This completes the proof  that the identity $\rho_s+\rho_n=\rho$, where both $\rho_s$ and $\rho_n$ are defined microscopically in an independent way, holds also in the absence of Galilean invariance.

{\em Conclusions}. In this letter we have derived explicit results for the normal density $\rho_n$  of a Bose-Einstein condensed gas with spin-orbit coupling.    The fact that the normal density does not vanish, as happens in usual Bose-Einstein condensed gases at zero temperature, is the dramatic consequence of the breaking of Galilean invariance, caused by the presence of spin-orbit coupling. We have shown, in particular,  that at the transition between the plane wave and the zero momentum phase, the effect is largest and the  normal density, associated with the flow along the direction of the momentum transferred by the Raman lasers,  coincides with the total density of the gas, with the consequent vanishing of the superfluid density. 
Our results set the stage for constructing a two-fluid model of a spin-orbit coupled Bose gas. Further important questions remain to be investigated, including the momentum of inertia of the gas in a trapped configuration, finite temperature effects and the investigation of second sound.

{\em Acknowledgement}. We would like to thank Anthony Leggett for discussions. Y.C. and S.Z. are supported by Hong Kong Research Grants Council, General Research Fund, HKU 17306414 and the Croucher Foundation under the Croucher Innovation Award. T.K. and S.Z. are supported by the CRF, HKUST3/CRF/13G. L.P and S.S. are supported by ERC
through the QGBE grant, by the QUIC grant of the Horizon2020
FET program and by Provincia Autonoma di
Trento.


\begin{thebibliography}{100}
\bibitem{Leggett2006} A. J. Leggett, {\em Quantum Liquids}, Oxford University Press, New York, 2006.


\bibitem{LS2003}  L. P. Pitaevskii and S. Stringari, {\em Bose-Einstein Condensation}, Clarendon, Oxford, 2003. {\em Bose-Einstein Condensation and Superfluidity}, Clarendon, Oxford, 2016.

\bibitem{Baym1969} G. Baym, in {\em Mathematical Methods in Solid State and Superfuid Theory}, edited by R.C. Clark and E.H. Derrick (Oliver and Boyd, Edinburgh, 1969).
\bibitem{GN1964} J. Gavoret and P. Nozi\'{e}res, Ann. Phys. {\bf 28}, 349 (1964).


\bibitem{Lin2011} Y.-J. Lin, K. Jim\'{e}nez-Garc\'{i}a,  and I. B. Spielman, Nature (London)\textbf{471}, 83 (2011).

\bibitem{JYZhang} J.- Y. Zhang,  et al., Phys. Rev. Lett. \textbf{109}, 115301 (2012).

\bibitem{JWPan} S. -C. Ji, J. -Y. Zhang,  L. Zhang, Z. -D. Du, W. Zheng, Y. -J. Deng, H. Zhai, S. Chen, J. -W. Pan, Nat. Phys. \textbf{10}, 314 (2014).

\bibitem{Zwierlein} L.W. Cheuk,  A.T. Sommer, Z. Hadzibabic, T. Yefsah, W.S. Bakr, M. W. Zwierlein, Phys. Rev. Lett. \textbf{109}, 095302 (2012).

\bibitem{JZhang1} P. Wang, Z.- Q. Yu, Z. Fu, J. Miao, L. Huang, S. Chai, H. Zhai, J. Zhang, Phys. Rev. Lett. \textbf{109}, 095301 (2012).

\bibitem{JZhang2} Z. Fu, L. Huang, Z. Meng, P. Wang, X.- J. Liu, H. Pu, H. Hu, J. Zhang, Phys. Rev. A \textbf{87}, 053619 (2013).

\bibitem{JZhang3} Z. Fu, L. Huang, Z. Meng, P. Wang, L. Zhang, S. Zhang, H. Zhai, P. Zhang, J. Zhang, Nat. Phys. \textbf{10}, 110 (2014).

\bibitem{Hamner} M.A. Khamehchi, Y. Zhang, C. Hamner, T. Busch, and P. Engels, Phys. Rev. A {\bf 90}, 063624 (2014)


\bibitem{Olson} A. J. Olson, S. -J. Wang, R. J. Niffenegger, C. -H. Li, C. H. Greene, and Y. P. Chen, Phys. Rev. A \textbf{90}, 013616 (2014).

\bibitem{SCJ} S.-C. Ji, L. Zhang, X.-T. Xu, Z. Wu, Y. Deng, S. Chen, and J.-W. Pan, Phys. Rev. Lett. {\bf 114}, 105301 (2015).

\bibitem{Engels2014} M. A. Khamehchi, Yongping Zhang, Chris Hamner, Thomas Busch, and Peter Engels
Phys. Rev. A {\bf 90}, 063624 (2014)

\bibitem{Zhu2012} Q. Zhu, C. Zhang, and B. Wu, Euro. Phys. Lett. 100, 50003 (2012)
\bibitem{Zheng2013} W. Zheng, Z.-Q. Yu, X. Cui, and H. Zhai, J. Phys. B: at. Mol. Opt. Phys. {\bf 46}, 134007 (2013).
\bibitem{Ozawa2013} Tomoki Ozawa, Lev P. Pitaevskii, and Sandro Stringari, Phys. Rev. A 87, 063610 (2013)



\bibitem{Yun2012b} Y. Li, G. I. Martone, S. Stringari, Euro. Phys. Lett. \textbf{99}, 56008 (2012).


\bibitem{Martone2012} G. I. Martone, Y. Li, L. P. Pitaevskii, S. Stringari, Phys. Rev. A \textbf{86}, 063621 (2012).

\bibitem{YunLi2015} Yun Li, Giovanni I. Martone, Sandro Stringari,  Annual Rev. of Cold Atoms and Molecules, World Scientific, Vol 3, 201 (2015)

\bibitem{Hui2015} H. Zhai, Reports on Progress in Physics {\bf 78} (2), 026001 (2015)

\bibitem{Ho2011} T.-L. Ho and S. Zhang, Phys. Rev. Lett. \textbf{107}, 150403 (2011).


\bibitem{Yun2012a} Y. Li, L.P. Pitaevskii, and S. Stringari, Phys. Rev. Lett. {\bf 108}, 225301 (2012).


\bibitem{Yun2013} Y. Li, G. I. Martone, L. P. Pitaevskii, S. Stringari, Phys. Rev. Lett. \textbf{110}, 235302 (2013).

\bibitem{Yu2014} Z.-Q. Yu, Phys. Rev. A {\bf 90}, 053608 (2014)

\bibitem{notesymm}The symmetrized notation, ensured by the term  ${\bf q} \to -{\bf q}$ in Eq.(\ref{defrhon}),  is required because the simultaneous breaking of parity and time reversal invariance leads to the inequality $S({\bf q},\omega)\neq S(-{\bf q},\omega)$~\cite{Martone2012}. 


\bibitem{note10} Results (\ref{mainresult})  and (\ref{c+c-}) are consistent with the expression $c_+c_-=\kappa^{-1}/(1+k_0^2\chi_M)$ derived in \cite{Martone2012} for the sound velocities.

\bibitem{Atkins1959} K.R. Atkins, Phys. Rev {\bf 113}, 962 (1959).

\bibitem{Fisher1973} M. E. Fisher, M. N. Barber and D. Jasnow, Phys. Rev. A {\bf 8}, 1111 (1973).


\bibitem{c+c-} As discussed in \cite{Martone2012} the sound velocities $c_+$ and $c_-$ are different in the plane wave phase if $G_2\ne 0$. In the zero momentum phase the identity $c_+=c_-$ always holds.

\end{thebibliography}
\end{document}